\documentstyle[11pt]{article}
\begin{document}
\begin{center}
\large{
The Faulty Assumptions of the Expanding-Universe Model vs. the Simple and Consistent Principles of a Flat-Universe Model \\
---- with Moving Pisa Tower Experiment which Tests General Relativity }
\normalsize\\
Jin He \\
Department of Physics, Huazhong University of Science and Technology, \\Wuhan, 430074, P.\,R.\,China\\
E-mail:mathnob@yahoo.com\\
\mbox{   }
\end{center}
\large {{\bf Abstract} } \normalsize

Einstein's personal life did not contradict the following facts.

Firstly, if Einstein stood on the ground, the ground gave him supporting force in balance of the Earth's gravity. Therefore, his cells suffered the force of pressure from each other from bottom to top. The cells of astronauts in International Space Station, however, suffer no such pressure. Therefore, the astronauts suffer a 22\% blood volume decrease or loss within 2 to 3 days. Muscle atrophy occurs at 5\% per week. Bone atrophy occurs at 1\% per month. Remember, our bodies were developed within the earth's gravity field, our muscles are needed to counter gravity. In space, we no longer need our muscles,so the brain starts to optimize the body by getting rid of what it doesn't need.

Secondly, Mach Principle presents the absolute universe. For example, when Einstein stood on the ground and relaxed, his arms fell down naturally. However, if he rotated his body then his arms were lifted up as the rotation was faster and faster. Mach principle is that the matter of the whole universe can affect local dynamic systems. That is, the matter of the whole universe sets up the local absolute reference frames.

However, Einstein's general theory of relativity is against the absolute reference frames of Mach Principle and can not explain the origin of life. Therefore, relativity is not a religion and Einstein can not be considered a god!
\\
keywords: Special Relativity -- General Relativity -- Gravitational Theory –- Hubble Law -- Cosmology
\\
\\

\section{ The Absolute Reference Frame of the Universe}
{\sf Human beings have many arrogant concepts. One of them is this:
people are willing to put themselves as the center of the universe
and thought that they did not belong to the nature and, instead,
they were the spiritual existence independent of the nature.

In fact, it is simple to verify that people are physically
originated. If people could find one event which breaks causality,
whether it is related to natural world or human beings itself, then
the assumption of spiritual origin of human would have a solid
foundation. We call such event ghost-event. Unfortunately, mankind
has not found such ghost-event that is contradictory to causality.

Einstein's theory of relativity is about high-speed motion of
massless particle or the interaction of huge masses. It is difficult
to be directly verified. But it tends to be developed into some
singular theories against causality. Time tunnel, time travel, black
hole, wormhole, the Big Bang and so on are such theories which try
to breach causality but are never verifiable.

If the entire universe forms the absolute reference frame then such
ghost-events have no basis. This is because everything in the
universe has the common standard of reference. Astronomical
observations do confirm the existence of the absolute reference
frame of the universe!

           The universe is full of materials most of which are
galaxies. Observations show that the materials are uniformly
distributed at large scales. They move with respect to each other.
Today's Big Bang cosmology is based on the assumption that the
motion have no absolute reference. If there were no absolute
reference
 then the universe would be unknowable and there would be no law in the
observational data about the large-scale universe. However, ten
years after Einstein had proposed his general relativity and
cosmology, Hubble discovered an important law about the universe,
called the Hubble redshift law. It says that distant galaxies issue
redshifted spectra which are proportional to their respective
distances from earth. Big Bang cosmology assumes that galaxies ran
away from us and the expansion led to the redshifts as runaway
trains give the static audience the sound waves of weaker
frequencies. However, if there were no convergent movement among the
distant galaxies then some galaxies would move towards the earth at
fantastic speeds. The speeds would be so large that they would
overwhelm the expansion and hit the earth directly. In that case,
the galaxies would issue blueshifted spectra rather than redshifted
ones. However, distant galaxies always present redshifted spectra
and the redshifts are proportional to their distances from Earth!

         Therefore, mainstream physicists and astrophysicists
admit that the materials in the universe form the absolute reference
frame. But they can not explain it. The existence of the absolute
reference frame, however, is one of many consistent conclusions made
by Dr. He's simple model of the universe.

Recent astronomical observations show that the cosmic microwave
background radiation has the privileged reference frame in which the
radiation is isotropic.
 This privileged frame is exactly the absolute reference frame of the universe.
Moreover, Earth moves with respect to the reference frame at the
speed of several hundred kilometers per second which matches with
the magnitude of the ``ether drift''. This is the precise
observation of the absolute reference frame of the universe. It is
the devastating blow to  the Big Bang cosmology.

       Common people have never heard of scientists talking about the
absolute reference frame. This is because it is against the tenet of
Einstein's general theory of relativity. Could the theory of
relativity explain why the absolute reference frame exists?
Similarly, how would the relativistic tenet explain the pattern of
galaxies? Therefore, scientists never openly talk about the patterns
of galaxies! But the pattern of galaxies is meaningful, and must be
explained by the matriarchal and patriarchal orders.

In fact, the Big Bang cosmology is a death universe without any
meaning: no reference frame, no origin of changes, no origin of
structures, no origin of time, no causality!

Einstein's general theory of relativity has no room for reference
frames.
  Dr. He % mmm
   made a little modification of general relativity by adding
the background reference frames to the theory. Such little
modification, however,  can explain the planetary distribution in
the solar system and explain the data of ``Gravity Probe B''
experiment performed by Stanford University and NASA (if the data
finally show that Einstein field equation is wrong). The following
chapters discuss the issues of solar system. These issues are
directly related to human environment and are part of the local
patriarchal order. }

\section{   The Grand Design of Universe }

\subsection{  Expression of constant density }
{\sf  Ideal universe is constant. Constant density means that all
 objects have constant velocity (no change with time).
 Velocity is usually defined to be the variance rate of
spatial distance with respect to time. The definition does not treat
time $t$ and space $x$ equally. Actually we can introduce any
parameter $p$, and calculate variance rate of time $t$ and distance
$x$ with respect to the parameter $p$: $T$ and $X$ respectively.
Careful readers will find in the following formulas that $T$ is the
variance rate of $t$ multiplied by the speed of light $c$ at current
cosmological time. Therefore, the expression of constant density is:
$$
            L = XX - TT
$$
The above expression is called Lagrange functional. The motion of
any object can be solved according to the Lagrange functional. This
method is known as the principle of optimization. Except human
activities, nature always obeys the principle of optimization!
According to the Lagrange functional of constant density, the reader
can prove that all objects have constant velocities, and the maximum
speed is the speed of light.

         In fact, space is three-dimensional while time is always
one-dimensional. Therefore, the above-said functional should be: $ L
= XX + YY + ZZ - TT$. However, for simplicity, we choose to ignore
the additional $Y$ and $Z$ terms.}
% 24.3
\subsection{  Real universe: changing density with time}
  {\sf       A universe of constant density is not real.
It is a dead one without vitality. Realistic one is that the
distribution of materials is spatially homogeneous at large-scales
but the density changes over time (aging). This changing universe
presents the force which has the similar effect as fluid pressure:
it exerts at any point in all directions. It is this pressure that
indicates an starting point of the universe. The universe itself has
a beginning like human beings!

         We will know that the simple assumption of aging density is
able to explain all basic astronomic observations.}

% 24.4
\subsection{  The Lagrange functional of the real universe }
   {\sf     For the motion of objects in the real universe, the
corresponding Lagrangian functional is different from the above one
(24.1) by  two factors, $A (t)$ and $B (t)$:
$$
            L = A (t) XX - B (t) TT
$$
These two factors are independent of spatial variables because the
density of materials at the large-scales is spatially uniform.
Therefore,
 Dr. He's % mmm
  model of the universe contains only two variables $A$ and
$B$.

      Do you want to learn more about the origin of the universe?
      Do you want to learn more
about the origin of earth structure? Please find the formulas of
particle motion based on the Lagrange functional of real universe
(24.2). If you have learned college mathematics seriously then it is
easy for you to derive the formulas. This is the basic skills for
you to understand the origin of natural structure.}
% 24.5
\subsection{  Cosmological redshift and Hubble redshift law }
  {\sf     Astronomical observations show that stars in distant galaxies
present atomic spectrum whose frequency is weaker than the one
observed on Earth. This resembles the phenomenon that the siren
frequency from moving-away train is weaker than the one from the
still train. This is called Kepler redshift of motion. However, do
you really believe that galaxies in the universe move away from us?
The universe is vast and the light traveling from one end of our
galaxy to the other end takes millions of years. As for the distant
galaxies, we simply can not observe their single stars, not to
mention the star motion on the sky.

       The redshift of the universe is actually the symbol of aging universe.
Big Bang cosmology has no reference frame but real universe has the
absolute reference frame. The spatial variables in the
above-mentioned formula (24.2) are the ones defined in the absolute
reference frame while the aging  universe defines the time variable.
For simplicity, we take the universe's current aging process to be
the standard time. The optimization principle proves (\cite{he5})
that the observation of atomic frequency spectrum depends only on
the coefficient $B (t)$ and the redshift of the spectrum requires $B
(t)$ increases with time $t$!

        Hubble discovered an important law known as the Hubble redshift
law: the distance of a galaxy from the earth is proportional to the
corresponding redshift of the galaxy.      The proportion constant
is called the Hubble constant. Calculation of the distance involves
the factor $A (t)$ in the formula (24.2). Therefore, Hubble redshift
law requires that the factor $A (t)$ be dependent on the other
factor $B (t)$.

       Dr. He's % mmm
        model of the universe contains only two variables $A (t)$ and $B (t)$.
 Cosmological Redshift requires that $B (t)$ increase with
time while the Hubble redshift law requires that $A (t)$ depend on
$B (t)$. It looks that
  Dr. He's % mmm
  model of the universe would fail. Only one
variable is left and its direction of monotonous change with time is
identified, however, we still have a lot of astronomical
observations to be explained by the model.}

% 24.6
\subsection{  ``Accelerating expansion'' of the universe }
  {\sf       Astronomical observation in 1998 shows that the Hubble constant
is not a constant, but an increasing variable with time!

         Big Bang cosmology is based on the general theory of relativity,
and general relativity is based on Newton's gravity. Gravity means
that objects move more and more  closer. Therefore, Big Bang
cosmology predicts that the expansion of the universe should be
slower and slower due to gravitation. That is, Big Bang cosmology
assumes that the universe should be expanding and the expansion
should slow down. However Observations show that the universe is at
``accelerating expansion''. The authorities never feel embarrassed:
they assume that the main component of the universe is the never
observable dark material which has negative energy, called dark
energy. The negative energy presents repulsive force
 so that the universe had ``accelerating expansion''.
Anyway, no common people can understand it.

       But
         Dr. He's % mmm
          model of aging universe directly indicates that the
Hubble constant is an increasing function with time if $B (t)$ is an
increasing function with time.}
% 24.7
\subsection{  The speed of light is not constant, but decreases with time! }
  {\sf      Big Bang cosmology is based on the general theory of relativity
and general relativity assumes that the laws of physics (including
physical constants) are the same at any time and any where. However,
astronomical observations show that the fine structure constant of
atomic physics changes with the evolution of the universe!

           The universe must have a violent start. In order to achieve the
uniform distribution of materials in the later time we have to
assume that the speed of light was close to infinity at the starting
point of the universe and then decreases. In other words, the speed
of light decreases with time. According to
  Dr. He's % mmm
   model of aging universe,
the decreasing speed of light is consistent with the requirement
that $B (t)$ increase with time. What a miracle!

       To overcome the above-said issue,  Big Bang theory assumes that the
starting explosion of the universe should be immediately followed by
a process of inflation. This process is not testable which was made
by some scientists to ``resolve'' the big problem and ``save'' the
big bang theory.}
% 24.8
\subsection{  The absolute reference frame of the
universe }
    {\sf   If there were no absolute reference frame of the universe then
cosmological law of the universe would not exist. However, the
universe is observable and it has laws. The first observed
cosmological law of the universe is the Hubble redshift law. It is
the fundamental evidence that the universe has the absolute
reference frame.

        Recent astronomical observations show that the cosmic microwave
background radiation has the privileged reference frame in
which  the radiation is isotropic.
 This privileged frame is exactly the absolute reference frame of the
universe. Moreover, the Earth moves with respect to the reference
frame at the speed of several hundred kilometers per second. This is
the precise observation of the absolute reference frame of the
universe. It is the devastating blow to
 the Big Bang cosmology.

        According to
          Dr. He's % mmm
           model, the universe is aging and the objects
in the universe (galaxies, for example) tend to be static with
respect to each other. That is, the motion with respect to each
other slows down and approaches the ultimate mutual stationary
positions. This mutual static process forms the universe's absolute
reference frame! With respect to the absolute reference frame, the
speed of any body in the universe is a decreasing function over
time. This is consistent with the requirement that $B (t)$ increase
with time. What a miracle too!}

% 24.9
\subsection{  Structure formation of the universe  }
     {\sf    Our universe is composed of stars. Stars are constantly
burning: turning massive particles into massless photons in order to
illuminate the universe's structure (including humans).

        As a result, the mass density of the universe is
decreasing.
Dr. He's % mmm
 model of the universe and quantum gravity point out
that mass density of the universe at large scale does decrease with
time, a fact consistent with the increasing function $ B (t)$! This
is really the miracle of miracles!}

% ch25

Full mathematical details of the the above cosmological model are given in the following Section.

\section{ A Flat-Universe Model Based on Simple and Consistent Principles }

{\it (i) A set of simple and basic principles about the large-scale universe.}
Our current knowledge of the universe is very limited and all models of the universe are mainly based on some assumptions. The only must-explained observational facts by any model are the Hubble law that all galactic spectral-line redshifts are proportional to corresponding galactic distances from us, and the `accelerated expansion`. The standard expanding-universe model follows the curved-spacetime assumption of general relativity (GR) and galactic redshifts are accordingly believed to be the Doppler-effect of the assumed galactic recession. However, curved spacetime is a faulty assumption. A correct model of the universe must be based on some absolute inertial frame (the flat background spacetime). The existence of such a frame is shown to be true from the following three basic principles:
(1) the universe has an isotropic but temporally inhomogeneous gravitational field; (2) the gravity is described by a Lagrangian which is the generalization to the proper distance of special relativity (the metric form of GR); (3) Hubble law is approximately true. These lead to varying light speed with time and give account of galactic redshifts and Hubble law (including `accelerated expansion`) as demonstrated in the following.

{\it (ii) Lagrangian, Lagrange$^,$s equation and its solution. }
The above set of principles uniquely determine the following Lagrangian which describes the isotropic gravitational field in the universe,
\begin{equation}
\begin{array}{ll}
\frac {1}{2}\grave {\bar s} ^2 &= L(x^0, x^i, \grave x ^0, \grave x ^i) \\
&= \frac{1}{2}B(\tilde  t )(\grave x ^0)^2- \frac{1}{2} A(\tilde  t )\sum ^{3}_{i=1} (\grave x ^i)^2   \\
&=\frac{1}{2} g_{\alpha \beta } \frac{dx^\alpha }{dp}  \frac{dx^\beta }{dp}
\end{array}
\end{equation}
where
\begin{equation}
g_{00}=B(\tilde  t )(>0), g_{11}=g_{22}=g_{33}= -A(\tilde  t )(<0), \mbox{ } g _{\alpha \beta } = 0 (\alpha \not= \beta).
\end{equation}
The canonical momentums conjugate to time $\tilde  t$ and space coordinates $x^i$ are respectively,
\begin{equation}
\begin{array}{l}
P_0 =\frac { \partial }{\partial \grave x^0} L=
B \frac{d\tilde  t}{dp}   \\
P_i =\frac { \partial }{\partial \grave x^i} L=-A\frac{dx^i}{dp}, \mbox{ } i=1,2,3.
\end{array}
\end{equation}
To find the Lagrange$^,$s equation, we need
\begin{equation}
\begin{array}{l}
\frac { \partial }{\partial x^0} L=\frac{1}{2} B^\prime  \left ( \frac{d\tilde  t}{dp} \right )^2- \frac{1}{2} A^\prime \sum^3_{i=1} \left ( \frac{dx^i}{dp} \right )^2 ,
  \\
\frac { \partial }{\partial x^i} L= 0, \mbox { } i=1,2,3, \\
\frac {d}{dp}\left ( \frac { \partial }{\partial \grave { \tilde  t}} L \right )
= B^\prime  \left ( \frac{d\tilde  t}{dp} \right )^2 + B \frac{d^2\tilde  t}{dp^2}
\end{array}
\end{equation}
where $A^\prime $ and $B^\prime $ are derivatives with time $\tilde t$:
\begin{equation}
  A^\prime =\frac {dA(\tilde t)}{d\tilde t} ,\mbox{ }
B^\prime =\frac {dB(\tilde t)}{d\tilde t} .
\end{equation}
The middle equation in (35) indicates that $x^i, i=1,2,3$ are cyclic coordinates. Therefore, the spatial components of the Lagrange`s equation are
\begin{equation}
P_i ={\rm constant } =- A(\tilde  t)\frac{dx^i}{dp}=, \mbox { } i=1,2,3.
\end{equation}
The temporal component is
\begin{equation}
\frac {d}{dp}\left ( \frac { \partial }{\partial \grave { \tilde  t}} L \right )-\frac { \partial }{\partial \tilde  t} L
=B \frac {d^2\tilde  t}{dp^2}+ \frac{1}{2} B^\prime  \left ( \frac{d\tilde  t}{dp} \right )^2 + \frac{1}{2} A^\prime \sum^3_{i=1} \left ( \frac{dx^i}{dp} \right )^2 =0.
\end{equation}
Combination with (37) gives
\begin{equation}
B \frac {d^2\tilde  t}{dp^2}+ \frac{1}{2} B^\prime  \left ( \frac{d\tilde  t}{dp} \right )^2 + \frac{A^\prime }{2A^2}( (P_1)^2+(P_2)^2+(P_3)^2 )=0.
\end{equation}
We define the constant (conservative) spatial momentum $P$ of the particle (e.\,g., a galaxy or a light crest from a galaxy),
\begin{equation}
 P^2 =(P_1)^2+(P_2)^2+(P_3)^2 .
\end{equation}
Finally the solution of the equation (39) is
\begin{equation}
\frac {dp}{d\tilde  t}= \sqrt{ \frac{ A(\tilde  t)B(\tilde  t)}{P^2+WA(\tilde  t) }}.
\end{equation}
where $W$ is another constant.
Substitution of the solution to the spatial ones (37) we finally have
the particle$^,$s motion in our universe
\begin{equation}
\frac {dx^i}{d\tilde  t}= -P_i \sqrt{ \frac{ B(\tilde  t)}{(P^2+WA(\tilde  t)) A(\tilde  t)}}.
\end{equation}
However, this is not the full story. Since our Lagrangian is proportional to the effective distance $(d\bar s/dp)^2$ and we deal with causal motion only, we always have $d\bar s^2 \geq 0$. Substitution of all our solutions into (32) gives
\begin{equation}
\frac {1}{2} \left( \frac{d\bar s}{dp}\right ) ^2 =L= \frac{1}{2}W={\rm constant }.
\end{equation}
This supports our previous assertion
\begin{equation}
\bar s \propto p.
\end{equation}
Causal motion implies that we always have
\begin{equation}
W\geq 0.
\end{equation}

{\it (iii) Varying light speed in the gravitational field of the universe. }
Because light has the maximum speed ($d\bar s^2=0$), we have $W=0$ for the motion of light. In its propagation direction we have
\begin{equation}
\frac {dx}{d\tilde  t}= \sqrt{ \frac{ B(\tilde  t)}{A(\tilde  t)}}.
\end{equation}
Currently the universe is at the time of
\begin{equation}
\tilde t=\tilde t_1 =ct_1.
\end{equation}
The current light speed is $c \simeq 3\times 10^8 {\rm m\,s}^{-1}$ which is used in the definition of $\tilde t $: $\tilde t = ct$. It is not wrong that we choose other light speed for the definition.

{\it (iv) Galactic redshift and Hubble law. }
Galactic redshift is the formula (12)
\begin{equation}
z=\frac{\nu _1 }{\nu _2} -1= \frac{ \sqrt{g _{00}( \tilde  t _1)} }{\sqrt{g _{00}( \tilde  t _2)}}-1
=\frac{ \sqrt{ B(\tilde  t_1 ) } }{\sqrt{B(\tilde  t _2) }} -1.
\end{equation}
We see that $B(\tilde  t )$ must be a monotonously increasing function with time for us to have galactic redshifts rather than blueshifts,
\begin{equation}
B(\tilde t) \uparrow .
\end{equation}
The distance $D$ between the two galaxies 1 (Milky Way) and 2 is given by the integral of the light travel formula (46)
\begin{equation}
D= \int ^{\tilde  t_1} _{\tilde  t _2}\frac {dx}{d\tilde  t}d\tilde  t= \int ^{\tilde  t_1} _{\tilde  t _2} \sqrt{ \frac{ B(\tilde  t)}{A(\tilde  t)}} d\tilde  t.
\end{equation}
The distance formula must have a redshift factor to give the Hubble law.
This indicates that $A(\tilde  t )$ depends on $B(\tilde  t )$. A simple and general model of the dependence is
\begin{equation}
A(\tilde  t )= \frac {B^{m+1}(\tilde  t)}{N^2B^{\prime 2} (\tilde  t) }
\end{equation}
where $m$ is a constant and $N(>0)$ is another constant whose unit is length.
Finally we have Hubble law,
\begin{equation}
\begin{array}{ll}
D&= \frac{2N}{m-2}\left (\frac{1}{\sqrt{ B(\tilde  t_2) }^{m-2} } -\frac{1}{\sqrt{ B(\tilde  t_1) }^{m-2} }\right )\\
&= \frac{2N}{m-2}\left (\frac{1}{\sqrt{ B(\tilde  t_2) } } -\frac{1}{\sqrt{ B(\tilde  t_1) } }\right )\left(\frac{1}{\sqrt{ B(\tilde  t_2) }^{m-3} } +\dots  \right ) \\
&=\frac{2Nz}{(m-2)\sqrt{ B(\tilde  t_1) }}\left(\frac{1}{\sqrt{ B(\tilde  t_2) }^{m-3} } +\dots  \right ) \\
&=\frac{cz}{H_0(\tilde  t_2, \tilde  t_1)}
\end{array}
\end{equation}
where the Hubble constant $H_0$ is
\begin{equation}
H_0=\frac{c(m-2)\sqrt{ B(\tilde  t_1) }}{2N}/\left(\frac{1}{\sqrt{ B(\tilde  t_2) }^{m-3} } +\dots  \right ).
\end{equation}
As a summary, I note that the redshift requires $B(\tilde t )$ be a monotonously increasing function of time and Hubble law requires $A$ be determined by the function $B$ (see (51)).
Therefore, the only one degree of freedom left is the function form of $B(\tilde t )$.

{\it (v) 'Accelerated Expanding$^,$ Universe.}
If $H_0$ depended only on $\tilde  t_1$, the current time, then Hubble law would be perfectly true.  However, it depends on the past time of the galaxy we observe,
\begin{equation}
H_0= H_0(\tilde  t_2, \tilde  t_1).
\end{equation}
If we assume
$$
m>3
$$
then Hubble constant $H_0$ is not constant and increases with the past time $\tilde  t_2$, of which the galaxy is observed.  This increase with time of $H_0$ is explained as the `accelerating expansion$^,$ of the universe. However, in my model, spacetime is flat (no expansion of curved spacetime) and the redshift is gravitational one which results from the evolution of the universe (mass density varies with time).
Because redshift requires increasing $B(\tilde  t)$, we see that `accelerating expansion$^,$ is consistent to galactic redshift.

{\it (vi) Infinite Light Speed and the Birth of the Universe.}
Positive and increasing quantity $B(\tilde  t)$ indicates a time $\tilde  t _0$,  when $B(\tilde  t _0)=0$. This is the starting time of the universe. We can choose $\tilde  t _0= 0$ to be the time of birth. Currently we do not know the exact physics at the hot birth. One thing is sure that light speed at the time must be infinite. Only infinite speed of communication could result in a later spatially homogeneous mass distribution in the infinite flat universe. This resolves the horizon and flatness problems due to birth. Infinite initial light speed indicates a decrease of light speed with time. Observation during the last decade does support the result of decreasing light-speed with time. The formula of light speed is (46). Therefore, decreasing light speed imposes further condition on the evolving factor $B(\tilde  t)$,
\begin{equation}
     2BB^{\prime \prime} \le mB^{\prime 2}.
\end{equation}

{\it (vii) Light Speed Constancy and the Death of the Universe.}
However, there is strong evidence that light speed is approximately constant during mature stage of the universe. Constant light speed with time means that $A(\tilde  t) $ and $B(\tilde  t)$ are the same
$$
  A(\tilde  t) \equiv B(\tilde  t) .
$$
They serve as the scaling factor. Perfect Hubble Redshift-distance linear law completely determines the scaling factor,
\begin{equation}
     \frac {1} {B(\tilde  t) } \equiv \frac {1} {A(\tilde  t) } = \frac {1} {B_0 } –- M (\tilde t - \tilde  t_0)
\end{equation}
where $M$ is a constant and $ B_0 = B(\tilde  t_0) $. This formula indicates a finite time $ \tilde  t_1$  when $ M(\tilde  t_1-\tilde  t_0  )= 1/B_0 $.  This is the ending time of the universe because the scaling factor reaches infinity. The possibility of a rebirth needs further investigation.

{\it (viii) The Absolute Inertial Frame of the Universe.}
Our calculation and results are reference-frames depended. For example, photon frequency is dependent on reference frames. Our results are meaningful only when single preferred inertial frame of the universe exists and the results are calculated with respect to the frame. The absolute frame is meaningful only when all components (e.g., galaxies) of the universe have convergent motion with respect to the frame. That is, all components slow down their speed of motion with respect to the frame. Since the nineteenth century, scientific report on the evidences of absolute inertial frame has never been stopped. Because of light speed constancy we have $  A(\tilde  t) \equiv B(\tilde  t) $ in the formula (42). We can see that the absolute speed of material particles (galaxies) does decrease with time, slowing-down motion with respect to the absolute inertial frame (note that $W > 0$ for material particles). Here we see that the existence of absolute inertial frame is once again the direct result of galactic redshift.

{\it (viiii) The Variance with Time of Matter Distribution in the Universe.}
Our Lagrangian is defined on flat spacetime and can be quantized according to the classical and covariant quantization procedure (He, 2006). Because the spatial distribution of matter in the universe is homogeneous, the resulting amplitude of the wave function must be proportional to the density of the distribution. Astronomical observation suggests that the density decreases with time especially during early universe. We can see that the amplitude does decrease with time if $B(\tilde  t) $ increases with time. That is, the astronomic observation is once again consistent to the result of galactic redshift.

\section{  All are the Change of Materials}
% 25.1
\subsection{ The essence of
  Dr. He's % mmm
   model of the universe  }
    {\sf       You have known
             Dr. He's % mmm
              model of the universe. We help you
better  appreciate it! The essence of
  Dr. He's % mmm
   model of the universe is that
everything is the physical change of materials (including human)!}
% 25.2
\subsection{  The nature of time  }
 {\sf     The nature of time  is also the change of materials. When two years
ago we knew from the internet that British physicist Julian Barbour
had the idea of ``there is no time but change'', we had a sudden
wake up in our life! In fact, Julian Barber put forward  the idea in
more than 40 years ago when he was a doctoral student in physics.
Like Einstein, he was mature at his early age. Therefore, he was
aware that if he insisted on scientific truth he would not be able
to have a career in a university. Einstein relativity has become a
religion, and the concept of time has been demonized as an
independent existence (the object with dynamic energy) instead of
the fact that time stands for the change of real materials!

       In order to continue his exploration of truth, Julian Barbour, after
graduation, bought a house in the country and raised his family by
translating into English the Russian physics journals in the former
Soviet Union.

       What a pity!}
% 25.3
\subsection{  Space is also the change of materials  }
   {\sf     According to
          Dr. He's % mmm
           model of the universe, there is no imagined
        space. Instead space is the real existence of material:
there are infinite materials in position. Do not imagine a boundary
of the universe. There is no border, there is only materials. The
universe is the unmeasurable vast! However, the relative motion
between materials slows down gradually and eventually all the
materials will be static with respect to each other. Such a process
towards the final static states defines the absolute reference frame
of the universe. Because the large-scale distribution of materials
is always uniform, this frame is the global flat inertial reference
frame. The universe is stable which can be relied upon! The universe
sets the standard for the measuring of human life. Anyone who
recognizes this is modest. Only those arrogant people try to
overcome the vast.}
% 25.4
\subsection{  Have you seen time and space?  }
   {\sf     You have seen the old clocks, but they are the change of materials: the change of spring.
You have seen the atomic clocks, but they are the change of
materials: atomic radiation. You have seen the natural clocks on
farmlands, but they are the change of materials: sunrise and sunset.
You have seen the age, but it is the change of materials: faces.

       You have seen the scale, but it is real: wood. You have seen
the micro-world, but it is real: protons and electrons. You have
seen the macro-world, but it is real: forests and stars.}
% 25.5
\subsection{  All are the change of materials: the orderly changes  }
      {\sf     When you face your wife, you think of beauty, but have
never thought of that she is the orderly and rational change. When
you face your husband, you think of able man, but have never thought
of that he is the orderly and rational change.

  You have missed appreciation of the most important things in
  life!}

\newpage
\pagenumbering{arabic}

\begin{center}
\large{
Moving Pisa Leaning-Tower Experiment and a Review of my Research Work }
\normalsize\\
\end{center}

For over 90 years, a theory has never been verified by any physical experiment but regarded as truth for the explanation of astronomic phenomena and even the basis of the model of the whole universe. This theory is Einstein`s general relativity (GR). Very fortunately, NASA and Stanford University designed the first physical experiment to test GR. This is the satellite borne experiment of Gravity Probe B (GP-B) [1]. After more than 40 years of preparation, the experiment was finished whose results, however, will be announced in December of 2007. On 14 April 2007, at the American Physical Society (APS) meeting in Jacksonville, FL, a preliminary result was released with the graph (see below) which shows that Einstein is wrong!

I call the graph Pre-graph (preliminary result graph). If Pre-graph is confirmed in the December, people will ask where Einstein is wrong? Scientists deify GR because its weak-gravity approximation is testified by astronomic observations of the solar system and a few extrasolar objects. GP-B experiment tests the same weak-gravity too but its results are very accurate. If Pre-graph is confirmed then we can say that general relativity (GR) is correct in the first order approximation and is wrong in the second order! The basic equation of GR is Einstein field equation which, therefore, is wrong too. This is hardly surprising because the equation contains great parts of speculation. Its basis is the concept of curved space-time which must be amended. Therefore, Newton`s gravitational law is the first-order result of gravitation, and the higher-order result must contain new law. GR does not include this new law and, therefore, does not generalize Newtonian theory.

My suggestion of the new law is rotational gravity. This means that the rotational Earth produces not only the normal Newtonian gravity but also the additional gravity due to its rotation. Below I will propose a very simple physical experiment to testify the new gravity. This is also the most inexpensive physical test of GR! The experiment is called Moving Pisa Leaning-Tower experiment. Very fortunately, through the efforts of the past seven years, I made contributions to most important areas of astrophysics and cosmology in line with astronomic observations. Therefore, I give a simple account of the proposal of the experiment and a review of my previous research work, starting with my simple analytic model of galaxy patterns.

1. Symmetry and Analytical Expression of Galaxy Patterns \\
Human race is in the age of information and no result of scientific research could be hidden. Google Earth lets people at home see every corner of the Earth surface while Google Sky lets people see most observed objects and galaxies on the sky. These heavenly galaxies have charming patterns which are approximations of corresponding mass distributions and are the results of gravitational interaction. But general relativity (GR), the mainstream theory of gravity, is powerless for their  recognition. This also suggests that GR has problem. From the end of 2000 to the end of 2005, I spent almost five years to identify the underlying symmetry and the corresponding analytical expression of galaxy patterns [2,3]. For example, the two-dimensional patterns of spiral galaxies are determined by the orthogonal nets of curves of exponential type. The symmetry is that the ratio of mass densities on two sides of a curve at any point is constant along the curve. This simple symmetry of exponential index is in line with all observational laws of spiral galaxies. Firstly, astronomical observations show that the optical density of spiral disks decreases exponentially in radial direction. My symmetry derives this law. Secondly, the arms of spiral galaxies are logarithmically curved. My symmetry derives this law too. Thirdly, regular spiral galaxies have only two types: normal and barred. My symmetry allows only the two solutions. Fourthly, barred galaxies show, more or less, a set of symmetric enhancements at the ends of the stellar bar, called ansae, or the `handles` of the bar. My symmetry presents the required ansae. Fifthly, my symmetry explains the three-dimensional patterns of elliptical galaxies. This is really a miracle.

2. Potential Application of Galaxy Pattern Symmetry on Earth Typhoon \\
Nowadays, earth`s atmospheric pollution, environmental destruction, human-induced surface temperature increases, and climate changes have continued to worsen. The number of typhoons increases with their power strengthening. Anyone reading the satellite photos of typhoons could not resist suggesting the strong similarity between typhoons and spiral galaxies. Since the dynamical expressions of galaxies are found, is it possible to use the expression to study typhoons? I think it is very possible. Once typhoons are generated, their power is strengthened by various factors. However, if we know their formation mechanism, we can kill them in their cradles. Unfortunately, my expression of galactic dynamics is against the theory of general relativity, and I have neither fund nor opportunity to study typhoons.

3. Galactic Dynamics and Rotation Curves \\
My symmetry principle of galactic patterns leads naturally to galactic dynamics, and explains galaxy rotation curves [2]. A galaxy rotation curve is the radial variance of statistically averaging rotational speeds of the stars or other celestial bodies in the galaxy. General relativity once again is powerless for their explanation. The current galactic dynamics is based on Newtonian theory of gravity. Because the optical density of spiral galaxy disks decreases exponentially with radius, the rotation curves of galaxies would decrease too in accordance with Newton`s theory if galaxies consist of mainly luminous objects. Actual rotation curves, however, are either constant or increase with galaxy radius a little bit. To make both Newtonian theory and general relativity right at galaxy scales, relativists assume the existence of large amount of dark matters which surround galaxies so ingeniously that constant rotation curves are maintained. However, my galactic dynamics, which is based on pattern symmetry, explains galaxy rotation curves without the needs of dark matters. This is definitely not a coincidence. Therefore, there must exist some new law of gravity. GR does not include the new law and, of course, does not generalize Newtonian theory. I proposed this new law, the rotational gravity.

4. Rotational Gravity and its Explanation to GP-B Preliminary Results (Pre-graph)\\
My recent paper [4] is the interpretation of GP-B preliminary results (Pre-graph). The paper presents a new law of gravitation, the rotational gravity. This means that the rotational Earth produces not only the normal Newtonian gravity but also the additional gravity due to its rotation. The gravity is similar to Newtonian gravity, but is anisotropic. In the direction which is perpendicular to rotational axis, rotational gravity is the largest. Rotational gravity can explain GP-B preliminary results (Pre-graph) which deviate from the prediction of GR. Further, it is possibly the basis for the explanation of galactic constant rotational curves.

5. Moving Pisa Leaning-Tower Experiment which Testifies Rotational Gravity and GR \\
If GP-B preliminary result (Pre-graph) is confirmed, then GR must be wrong. My explanation is that GR does not include rotational gravity. In fact, we can make a very simple experiment to testify both GR and rotational gravity. This experiment is called Moving Pisa Leaning-Tower experiment. To obtain Earth mass, we usually measure gravitational acceleration on Earth`s surface. To measure the acceleration, we do what Galileo did by letting some objects freely fall from Pisa leaning-tower. Galileo found out that all objects of different masses hit the ground in the same time. That is, they have the same gravitational acceleration. To measure the exact value of this acceleration, we need only record the whole time interval of freely falling. However, Earth is always rotating and all our measurement is made in the non-inertial reference frame of rotation. The measured acceleration is the rotational-frame acceleration, and the corresponding mass is the rotational-frame earth mass. To get the inertial-frame acceleration, people usually add centrifugal acceleration to the rotational-frame one, according to Newtonian theory.

The question is: can we build a true inertial reference frame of Earth in the meaning of Newton? The answer is yes. If we allow the Pisa leaning-tower move uniformly at the speed of 1286.57 km per hour in western direction (actually we let the experimental apparatus move), then the movement of leaning tower cancels out earth rotation, and the Pisa leaning-tower is the true inertial reference frame of Earth in the meaning of Newton. The gravitational acceleration measured by the moving Pisa tower is the inertial-frame acceleration of earth, and the corresponding mass is the inertial-frame earth mass. This is called moving Pisa leaning-tower experiment. Is this inertial-frame acceleration equal to the one obtained indirectly by static tower? The answer given by Newtonian theory and GR is that they are approximately equal with immeasurable difference. However, if rotational gravity exists, the difference is measurable in the same way that GP-B preliminary result (Pre-graph) differs from GR prediction by measurable amount. Pre-graph indicates that earth rotational gravity is smaller than Newton gravity by a factor of about 207. According to moving Pisa leaning-tower, Earth is rotating, rotating mass produces rotational gravity, and every object has additional weight. If the additional weight corresponds to exactly the additional geodetic effect seen on Pre-graph and confirmed in the coming December, then rotational gravity is verified.

6. Significance of Rotational Gravity \\
Theoretical significance: Rotational gravity is similar to Newtonian gravity, but is anisotropic. In the direction which is perpendicular to rotational axis, rotational gravity is the largest. This can explain many astronomical phenomena. From our solar system to galaxies, there are many kinds of planar mass distributions. Their centers are usually objects of large rotational masses, and the rotational axes are approximately perpendicular to the planes. This can be easily explained by rotational gravity. Perpendicular to the rotation axis, gravity is larger and attracts more materials. Therefore, the moon moves on Earth`s equatorial plane. Earth atmosphere gathers more air in the equatorial direction. Natural satellites and planetary rings are usually on the equator planes of corresponding planets. All planets move approximately on solar equatorial plane. However, Newtonian theory and general relativity (GR) can not explain these phenomena!

Economic significance: Earth`s resources are limited. Today, more and more people use dynamical transportation. These transportation tools consume energy because they must overcome Earth`s gravity to do work. However, rotational gravity can be used to offset partially Earth`s gravity. For instance, on the top of any aircraft can be put a high-speed rotational mass. The rotational gravity produced by the mass can partially offset the earth gravity from beneath the aircraft. Because rotational gravity is squarely proportional to the rotational speed, rotational gravity can be comparable to aircraft weight if the speed is large enough.

7. General Relativity View of Space-time vs. my View of Space-time\\
General relativity view of space-time: GR is a theory of gravitation and the theory is very simple: in one sentence, gravity is curved space-time. Therefore, GR assumes that space-time is matter, a matter without mass and energy. In order to determine this strange matter, Einstein and his mathematician friends invented Einstein field equation to determine the strange matter. One side of the equation is the curvature of curved space-time while the other is the energy-momentum of real matters. It is straightforward to show that the equation is wrong. Einstein considered space-time to be physical. Therefore, curved space-time must have non-trivial topology as curved paper must have uneven shape. However, Einstein field equation has not the slightest relationship with topology. In other words, Einstein`s field equation can not determine curved space-time which Einstein sought. Einstein in his life did not find any other equation to determine his curved space-time. Therefore, Einstein`s space-time is independent of physical matters. The quantum mechanics of 20th century has given us a lesson: any physics concept independent of physical matters is usually wrong!

My view of space-time: My view of space-time is based on true matter. Space is the extension of true matter while time is defined to be the real changes of real matters. Newton`s view of space-time is also built on real matters but he overlook the truth that causal relation has limited propagation speed. Therefore, Newton has the concept of absolute time. However, the values of my time and distance measurements are dependent on reference frames. Newton`s inertial reference frame is real moving body. For example, Earth`s inertial reference frame is the hypothetical Earth without its rotation. In fact, Earth`s inertial frame is not really a inertial frame because any test particle with respect to the frame is subject to Earth`s gravity. In real inertial frame, any test particle is either static or moving straightly at constant speed with respect to the frame. However, with further assumption that Earth mass were zero, the Newton inertial reference frame is my inertial frame of earth. In my inertial frame, any test particle is really static or moving straightly at constant speed with respect to the frame. Similarly, Newton`s solar inertial reference frame is my inertial frame of sun if we assume sun had zero mass. In the inertial frame of sun, any test particle is either static or moving straightly at constant speed with respect to the frame because solar mass were zero. Similarly, the Milky Way inertial reference frame of Newton is my inertial frame if we assume that Milky Way had zero mass. Therefore, my inertial reference frame is truly flat space-time! Then, how to describe Earth`s gravity by using the inertial reference frame of earth? According to classical mechanics, flat space-time description of gravity is a Lagrangian form. If there were no earth rotation, the Lagrangian form of earth gravity is a Schwarzschild metric form of general relativity. Because GR has no global inertial frame and curved space-time has infinite sets of mathematical coordinates which have no real meaning, Scharzschild metric has various expressions corresponding to the coordinate systems, such as the standard expression, isotropic expression, and so on. My theory of space-time always uses a flat inertial reference frame to describe gravity. For example, Earth`s gravity is expressed by the flat earth inertial frame. Therefore, a specific expression of Scharzschild metric is the Lagrangian form of earth gravity in my theory. Which is the expression? The answer is determined by experiments. I do not know the answer by now.

At this point, you know that the metric form of general relativity which describes curved space-time is taken to be the Lagrangian form of gravity on flat space-time in my theory. It may seem that I am crazy. In fact, I am not crazy. If you have an objective view of GR and calm down to listen to my explanation, you will be able to understand me. In actual mathematical calculations, my Lagrangian form is the metric form which describes curved space-time, and most formulas of GR are carried over to my theory. However, my space-time is always curved locally with respect to a global flat space-time. For example, the curved space-time due to earth`s mass is locally curved with respect to the flat inertial frame of earth. Similarly, the curved space-time due to the mass of sun is locally curved with respect to the global flat inertial frame of sun. We now understand that my theory of gravity is formally the general theory of relativity if we accept one assumption that the whole universe itself constitutes an inertial reference system. This is the flat and absolute inertial system which sets the standard for all changes of matters in the universe. I proved that this unique inertial system does exist and it can explain astronomical observations without contradictions and without the needs of dark matters and dark energy. Because the mass distribution of the universe is locally uneven and there exists local hierarchical structure, the corresponding inertial reference frames constitute a hierarchical structure too. The values of time and distance measurements are dependent on reference frames, giving us a mathematical sense of curved space-time, and most formulations of general relativity can be carried over formally to my theory of gravity.

In short, the major differences between general relativity view of space-time and my view of space-time are as follows. Firstly, the whole universe itself constitutes a flat background reference frame, and the local apparent structure of curved space-time is with respect to the flat background. Secondly, space-time, which is not true substances, is the mathematical impression of measurements with respect to inertial reference frames. Curved space-time, which is not true substances too, is the mathematical impression of measurement transformations between different reference frames. Thirdly, time basically does not exist. Anything measured is the   changes of real matters, and the tool which performs measurement is itself a changing material. For example, the precise tool of measurement is generally the employment of electromagnetic waves. Time is the illusion of physical changes, and the evolving universe provides the most basic time. Fourthly, Einstein field equation is wrong which does not include rotational gravity.

8. Quantum Gravity and the Law of Orbital Radii of Planets \\
The classic theory of gravity which is not quantized can only describe the movement of single particle. Therefore, classical theory can not be used to describe a system of many bodies. One such system is the solar system. General relativity is still not quantized and, of course, can not be used to describe the distribution of planets in solar system. Any quantization scheme of a force must have an independent background space-time. This is because quantization must be based on an independent causal relation. Causal relation is in fact the background space-time. You can quantize a force but you can not quantize its causal relation. However, Einstein thought that gravity is the background space-time and background space-time is the gravity. This is why general relativity can not be quantized over 90 years. Over 90 years, how much money is consumed in the game of quantizing general relativity?

However, my theory of gravity always has a flat inertial frame as its background. For example, Earth`s gravity is expressed by the flat earth inertial frame and sun`s gravity is expressed by the flat inertial frame of sun. If you must use the mathematical concept of curved space-time, then the curved space-time due to earth`s mass is locally curved with respect to the flat inertial frame of earth, and the curved space-time due to the mass of sun is locally curved with respect to the global flat inertial frame of sun. Therefore, my theory of gravity can always be quantized: quantization with respect to the background flat spacetime. To this end, I only employ the classical scheme of covariant quantization. I do not introduce any new constant. My results on the quantization of sun`s gravity do explain the distribution of planets in solar system [5]. This is really a miracle!

9. My Very Simple Model of the Universe [6]\\
Since Einstein`s curved space-time of pure mathematical meaning does not exist, the universe must be flat. Will a flat universe made of physical matters be consistent and give results in line with astronomical observations?

First, a flat universe should provide the unique and eternal reference frame. A universe without any reference frame is chaos, and no common standard is available for the measurement of any kind of changes. With a physical universe, how is the unique and eternal reference frame achieved? Newton's definition of inertial frames is not applicable because objects of either static or uniform linear motion can not define a reference frame for the whole universe. The only possibility is that all objects in the universe tend to be static with respect to each other. Such decelerating movement also suggests an evolving universe.

Second, the velocity of any object is not infinite. In any stage of the universe, particles of the largest velocity have zero mass. This condition and the one of uniform and isotropic universe require that the Lagrange form of particle motion in the universe depend on only two variables. Third, Hubble law requires that one variable be dependent on the other. Therefore, a flat universe has only one variable for its description. I call the variable cosmic one. The variance of the cosmic variable with the evolution time of the universe is either an increase or a decrease. The choice of increase (or decrease) must comply with all observational laws and there should be no contradiction! Fourth, flat universe has no expansion, and the cosmic redshift is gravitational one. Redshifts require that the change of cosmic variable with evolution time be increase. Fifth, astronomical observations show that the Hubble constant is not constant, but increases with the evolution time. This corresponds to the condition that the change of the cosmic variable is increase, a fact consistent to gravitational redshift. Sixth, the uniformity of the universe and the existence of horizon require that the change of photon speed with evolution time of the universe be decrease. This corresponds to the condition that the change of the cosmic variable is increase, a fact consistent to gravitational redshift once again. Seventh, a flat universe requires the unique and eternal reference frame. This corresponds to the condition that the change of the cosmic variable is increase, a fact consistent to gravitational redshift once again. No contradiction. Eighth, the average mass density of the universe decreases with the evolution time of the universe. This corresponds to the condition that the change of the cosmic variable is increase, a fact consistent to gravitational redshift once more. This is really a miracle! Should we believe in Big Bang theory?

10. Social Significance of the Recognition that Time does not Exist \\
Since the universe is flat, time which is simple but mysterious, does not exist. All are the changes of real matters. The evolution of the universe defines the absolute and eternal time. Local changes of matters have the evolution of the universe as a background. For example, the local space-time which is curved by a giant mass must be balanced by the flatness of the whole universe. Therefore, those kinds of extreme concepts of curved space-time, such as time machine, time tunnel, time reversal, space-time wormhole, black holes, Big Bang, etc. simply do not exist.

We should admit that many of extreme human motivation and behavior are due to human`s mysterious fear of life and the universe. In essence, they are the results of human`s hallucinations of the monster, i.e., time. The final rescue of humans is to eliminate their mysterious fear of life and the universe, to seek their healthy changes, to avoid their destructive activities induced by their hallucinations, and finally to lead to the harmonic existence in physical nature.

To test my theory, we wait for two more months for the final results of GP-B experiment, or we can perform the Moving Pisa Leaning-Tower experiment! \\
\\
$[1]$   http://www.einstein.stanford.edu \\
$[2]$   http://www.arxiv.org/abs/astro-ph/0510535 \\
$[3]$   http://www.arxiv.org/abs/astro-ph/0510536 \\
$[4]$   http://www.arxiv.org/abs/astro-ph/0604084 \\
$[5]$   http://www.arxiv.org/abs/astro-ph/0604084v4   \\
$[6]$   http://www.arxiv.org/abs/astro-ph/0605213v4

\end{document}